\documentclass[journal,twoside,a4paper]{IEEEtran}

\usepackage{color}
\usepackage{amssymb}
\usepackage{rotating}
\usepackage[T1]{fontenc}
\usepackage{lscape}
\usepackage{dblfloatfix}
\usepackage{xcolor}
\usepackage{subcaption}
\usepackage[ruled,vlined]{algorithm2e}
\usepackage[autostyle]{csquotes}
\usepackage{tikz}
\usepackage{acro}
\usepackage{siunitx}
\usepackage{enumitem}
\usepackage{multirow}
\usepackage{cite}
\usepackage{hyperref}
\usepackage{amsmath,amssymb,amsfonts}
\usepackage{tikz}
\usepackage{float}
\usepackage{algorithmic}
\usepackage{graphicx}
\usepackage{textcomp}
\usepackage{xcolor}
\usepackage{url}

\usepackage{comment}
\def\BibTeX{{\rm B\kern-.05em{\sc i\kern-.025em b}\kern-.08em
    T\kern-.1667em\lower.7ex\hbox{E}\kern-.125emX}}

\ifCLASSINFOpdf
\else
 
\fi

\usepackage[a4paper, total={180mm,257mm}]{geometry}

\begin{document}

%\title{Perspectives in hybrid edge-cloud computing: a case study on energy efficiency in buildings}
%\title{A hybrid tow-stage edge-based anomaly detection of building energy consumption}

%\title{Anomaly-Net: Deep Anomaly Detection of Building Energy Consumption Using Time-Series Imaging}

\title{Quadrotor Experimental Dynamic Identification with Comprehensive NARX Neural Networks}


%
\author{\IEEEauthorblockN{Khaled Telli\IEEEauthorrefmark{1},
Okba Kraa\IEEEauthorrefmark{1}, 
Yassine Himeur\IEEEauthorrefmark{7}, 
Mohamed Boumehraz\IEEEauthorrefmark{1},
Shadi Atalla\IEEEauthorrefmark{7},
Wathiq Mansoor\IEEEauthorrefmark{7},
Abdelmalik Ouamane\IEEEauthorrefmark{3} 
}\\
\IEEEauthorblockA{\IEEEauthorrefmark{1}
Energy-Systems-Modelling Laboratory (MSE), University of Biskra, Biskra, Algeria}\\
\IEEEauthorblockA{\IEEEauthorrefmark{3} LI3C Laboratory, University of Biskra, Biskra, Algeria}\\
\IEEEauthorblockA{\IEEEauthorrefmark{7}College of Engineering and Information Technology, University of Dubai, Dubai, UAE}\\
}


\maketitle

\begin{abstract}
This research paper delves into the field of quadrotor dynamics, which are famous by their nonlinearity, under-actuation, and multivariable nature. Due to the critical need for precise modeling and control in this context we explore the capabilities of NARX (Nonlinear AutoRegressive with eXogenous inputs) Neural Networks (NN). These networks are employed for comprehensive and accurate modeling of quadrotor behaviors, take advantage of their ability to capture the hided dynamics. Our research encompasses a rigorous experimental setup, including the use of PRBS (Pseudo-random binary sequence) signals for excitation, to validate the efficacy of NARX-NN in predicting and controlling quadrotor dynamics. The results reveal exceptional accuracy, with fit percentages exceeding 99\% on both estimation and validation data. Moreover, we identified the quadrotor dynamics using different NARX NN structures, including the NARX model with a sigmoid NN, NARX feedforward NN, and cascade NN. In summary, our study positions NARX-NN as a transformative tool for quadrotor applications, ranging from autonomous navigation to aerial robotics, thanks to their accurate and comprehensive modeling capabilities.
\end{abstract}


\IEEEpeerreviewmaketitle

\section{Introduction}
Quadrotors, a type of Unmanned Aerial Vehicle (UAV) system, have garnered significant attention in recent years due to their versatility and potential applications in various fields, including surveillance \cite{nwaogu2023application,chouchane2023improving, 10287318}, agriculture \cite{schad2023opportunities}, search and rescue \cite{zhu2023automatic,himeur2022using}, localization \cite{sulaiman2023radio} and aerial robotics \cite{1}. Their unique ability to hover, navigate complex environments, and execute tasks autonomously has established them as invaluable tools in industry and academia. However, realizing the full potential of quadrotors necessitates a comprehensive understanding of their dynamic behavior, precise modeling, and robust control \cite{16}.
The dynamics of quadrotors are characterized by their pronounced non-linearity, under-actuation, and multi-variable nature \cite{2}. These inherent complexities pose challenges to researchers and engineers who aim to develop effective control strategies and navigation systems. Accurate modeling of quadrotor dynamics is pivotal for ensuring stability, control, and optimal performance.
Traditional quadrotor modeling approaches often resort to linearization techniques, simplifying the intricate dynamics for control design \cite{3}. While these methods provide some insight, they frequently fail to capture the nuances of real-world quadrotor behavior. To address this limitation, we explore Neural Networks (NN), a subset of machine learning (ML) models inspired by the structure and function of the human brain \cite{himeur2023ai}.

NNs, encompassing a variety of types including convolutional neural networks (CNN) \cite{4} and RNN \cite{5}, have showcased exceptional capabilities in discerning complex, nonlinear relationships within data \cite{himeur2022deep,copiaco2023innovative,himeur2023face}. They are adept at image recognition, natural language processing, and time-series prediction \cite{daradkeh2023lifelong}. Lately, there has been a surge in research interest towards their potential in modeling and controlling intricate dynamic systems \cite{6}.
In this paper, we embark on a thorough analysis of the application of NARX-NN for precise quadrotor dynamic modeling and system identification. Our goal is to harness the capabilities of NN to unravel intricate dependencies within quadrotor dynamics, enhancing our prediction and control of their behavior. We present a study that evaluates the performance of NARX networks in comparison to other NN architectures and validate our models with extensive experiments.
Through our research, we aspire to augment the existing knowledge on quadrotor dynamics and further the advancements in aerial robotics by offering innovative insights into accurate modeling and control strategies.

\section{Dynamic Modeling And Hardware Configuration}
\subsection{Quadrotor Dynamic Modelling}
\begin{figure}[htbp]
\centerline{\includegraphics [width=0.44\textwidth, height=0.15\textheight]{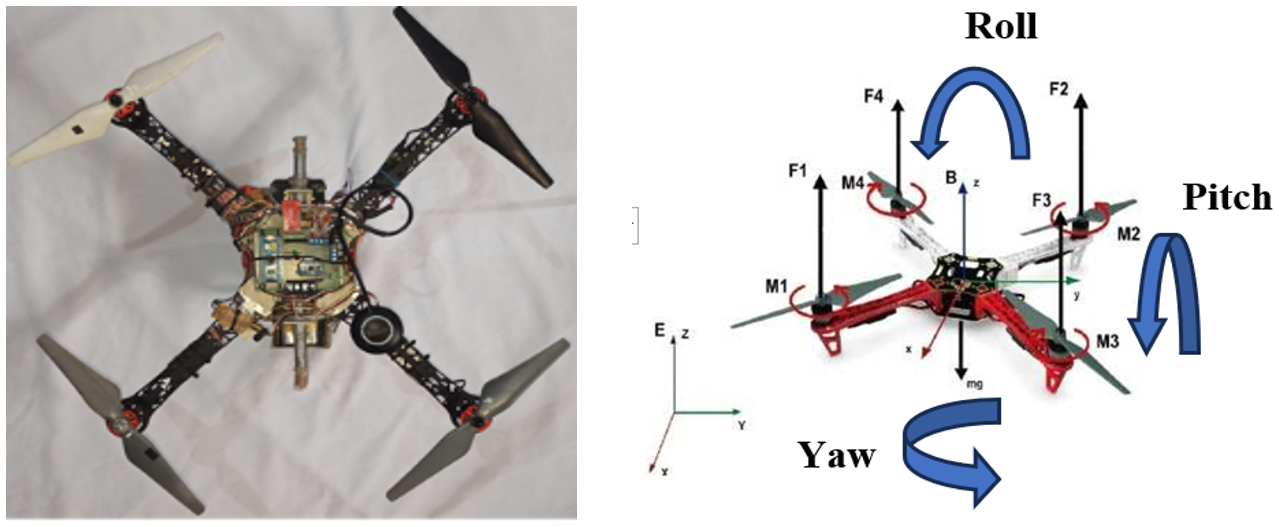}}
\caption{Used Quadrotor and applied.}
\label{fig1}
\end{figure}

The dynamic behavior of the quadrotor is distinguished by its pronounced nonlinearity, under-actuation, and multivariable nature. Despite having only four actuators, the quadrotor exhibits six degrees of freedom (DOF), as depicted in Fig. \ref{fig1}. The "E" reference frame represents the fixed-earth reference, while the "B" frame represents the body's inertial reference. This "B" frame is affixed to the gravity center of the quadrotor. The frame and blades of the quadrotor are assumed to be rigid and symmetric. The dynamical equations governing the behavior of the quadrotor are formulated as follows \cite{3}.

Fluctuations in rotor speeds influence the thrust forces, leading to various motions. Consequently, vertical movement arises from proportional increases or decreases in propeller speeds. Adjusting for roll rotation requires changing the velocity of the lateral motors (motors 3 and 2). In contrast, pitch rotation results from varying the speed of the rear motors (motors 2 and 4). Yaw rotation occurs due to differing speeds between the two pairs of propellers. It is vital to execute these maneuvers while maintaining consistent total thrust to preserve a steady altitude. The quadrotor's dynamics are characterized by instability, nonlinearity, multiple variables, and underactuation. Even with only four actuators, the quadrotor possesses six degrees of freedom.

\subsection{Quadrotor Hardware}
The quadrotor platform we use features an X-shaped, rigid, and symmetric body equipped with four E-MAX XA-2212 brushless electric motors. These motors can reach a peak speed of 920 RPM per volt and operate within a 7-12.6V voltage range. The quadrotor comes with 10x4.5-inch propellers and four SIMONK-branded electronic speed controllers (ESCs). These ESCs can handle a current load of 40A and require an input power of 5V for pulse width modulation. Flight control is powered by a lithium polymer (Li-ion) rechargeable battery, rated at 3S or 11.1 volts, with a 4500 mAh capacity. The STM32 blue pill microcontroller, which features a 32-bit ARM processor running at 72 megahertz, manages control. This controller offers 512 k-bytes of flash memory and 64 K-bytes of SRAM memory. The MPU-6050 Inertial Measurement Unit (IMU), incorporating three gyroscopes and three accelerometers, records the quadrotor's angular acceleration and velocity along the x, y, and z axes. The MPU-6050 communicates with the CPU via the I2C protocol. The quadrotor's altitude is ascertained by the MS5611 barometer. For remote control and communication, we use the AT9S R9DS system, which provides nine channels operating at 2.4GHz with PWM pulse durations between 1 to 2 ms. A 5 to 3.3V logic-level converter is utilized due to the STM32's 3.3V logic voltage level difference. The quadrotor's total weight is 1.05 kg, and calculations are performed at 250Hz (0.004 seconds). Fig. \ref{fig1} illustrates the experimental quadrotor and its hardware setup.

%\vspace{-0mm}
\section{Neural Network (NN)}
NNs are a class of ML models inspired by the structure and functionality of the human brain. They consist of interconnected nodes or neurons organized into several layers. These networks can learn patterns, relationships, and representations from data through a process known as training. As a component of artificial intelligence (AI), NNs have found widespread applications in UAV fields, including image and speech recognition, natural language processing, autonomous systems, and more, thanks to their capacity to tackle complex tasks by processing and learning from extensive datasets \cite{1,7}.

\subsection{Key NN Categories}
Over the years of studying NNs, research has led to the development of various classes and structures. Some of the commonly recognized classes of NNs, with their main features and applications, include \cite{8}:

\subsection{Feedforward Neural Network (FNNs) / Multi-Layer Perceptrons (MLPs)}
These are structured with three primary layers: input, hidden, and output. In this architecture, neurons—called nodes—within each layer form complete connections with the neurons in adjacent layers. This facilitates the unidirectional flow of data, from input to output, which typifies the connection pattern of FNNs. They are widely applied in a range of fields, especially in supervised learning tasks such as classification and regression. Their use is prominent in facial recognition, natural language processing (NLP), computer vision, and more.

\subsection{Convolutional Neural Network (CNNs)}
CNNs feature a unique architecture comprising several layers: convolutional layers, pooling layers, and fully connected layers. Within the CNN framework, convolutional layers are critical, applying filters to input data, which allows the capture of intricate spatial patterns. Information flows through a sequence of convolutional and pooling layers before reaching the fully connected layers, outlining the data's journey within the network. Designed primarily for tasks associated with image and grid-like data, such as image classification and object detection, CNNs also find utility in other domains like natural language processing (NLP) and optical character recognition (OCR), highlighting their adaptability and broad application spectrum.

\subsection{Recurrent Neural Network (RNNs)}
RNNs exhibit a unique structural arrangement wherein neurons are interconnected in a directed cycle. A defining feature of RNNs is their possession of a hidden state that retains information from previous time steps, contributing to their capability for processing sequences. The connection pattern in RNNs forms a loop, directing a flow of data in which each neuron receives input from the previous time step and produces output for the current one. This promotes an inherent temporal understanding within the network. RNNs excel in handling sequential data tasks, with applications spanning diverse domains like natural language processing and modeling and predicting time-series data. Their adaptability is evident in use cases such as text-to-speech conversion, sales forecasting, and stock market predictions, emphasizing their value in capturing temporal dependencies and making predictions based on historical context.

Other NN structures include Transformers, Autoencoders, Long Short-Term Memory (LSTM) Networks, Gated Recurrent Unit (GRU) Networks, Generative Adversarial Networks (GANs), Temporal Convolutional Network (TCN), Self-Organizing Maps (SOMs), Echo State Networks (ESNs), Modular NNs, Spiking NNs (SNNs), Attention Mechanisms, Siamese Networks, Capsule Networks, Neural Turing Machines, and more. However, the exact architecture of each network can vary based on specific implementation and task requirements. These are just a few of the many classes of NNs. Each class has its own set of variations and improvements, and the NN field is ever-evolving, with new architectures and techniques emerging regularly.

\subsection{NN For Dynamic Identification}
For dynamic system identification, many architectures of NN are used for this context. RNNs and variants of RNNs are frequently used. These networks are well-suited for modeling time-series data and dynamic systems because they can maintain a hidden state that captures temporal dependencies in the data. Here are some types of NN commonly used for dynamic system identification \cite{9,12}. Vanilla RNNs are the most basic type of RNNs. They have a simple recurrent structure that allows them to capture sequential information in the input data. However, they may suffer from the vanishing gradient problem, making it challenging to capture long-term dependencies. LSTM Network is a type of RNN designed to address the vanishing gradient problem. They have a more complex architecture with gates that control the flow of information, making them capable of learning and remembering long-term dependencies in time-series data. LSTMs are often used for dynamic system identification tasks due to their effectiveness. GRU Network is similar to LSTMs but has a simpler architecture with fewer gates. They are computationally less expensive than LSTMs and effective for modeling sequential data, including dynamic systems. ESNs are a specialized type of recurrent NN designed specifically for dynamic system identification. They consist of a large random recurrent hidden layer and a trainable output layer. ESNs have been used successfully in various applications to model and predict complex dynamic systems' behavior. Transformer-Based Models, Transformers were initially designed for natural language processing tasks. They have also been adapted for time-series data and dynamic system identification. Variants like the TCN and the Transformer-based models with positional encodings have been used to model temporal dependencies in sequential data. Hybrid models, in some cases, researchers combine traditional system identification methods with NN to create hybrid models. These models can take advantage of both the data-driven power of NN and the analytical techniques of classical system identification.
The choice of the NN architecture depends on the specific characteristics of the dynamic system and the nature of the available data. Researchers often experiment with different architectures to find the one that best fits the problem. Additionally, the training and validation of these networks may involve various techniques, such as hyperparameter tuning and optimization algorithms, to achieve the desired accuracy in identifying and predicting dynamic systems.

\section{NARX NN Structure}
The fundamental mathematical representation of NARX within a time domain system \cite{13,14} is outlined as follows. Consider a discrete-time system with a $k-th$ input-output sample $u(k)$ and $y(k)$, respectively, where $k$ represents the discrete time index. The NARX model aims to capture the relationship between past input-output pairs and current/future output values in a nonlinear fashion $f (.)$ which can include both linear and nonlinear sub-functions. The general NARX equation can be expressed as: 
%\vspace{-3mm}

\begin{equation}
\label{eq1}
\begin{aligned}
& y(k)=f(y(k-1), y(k-2), \ldots y(k-r), 
u(k- \\
& 1), u(k-2) \ldots, u(k-p))+e(k)
\end{aligned}
\end{equation}

\begin{figure}[htbp]
\centerline{\includegraphics[width=0.48\textwidth, height=0.16\textheight]{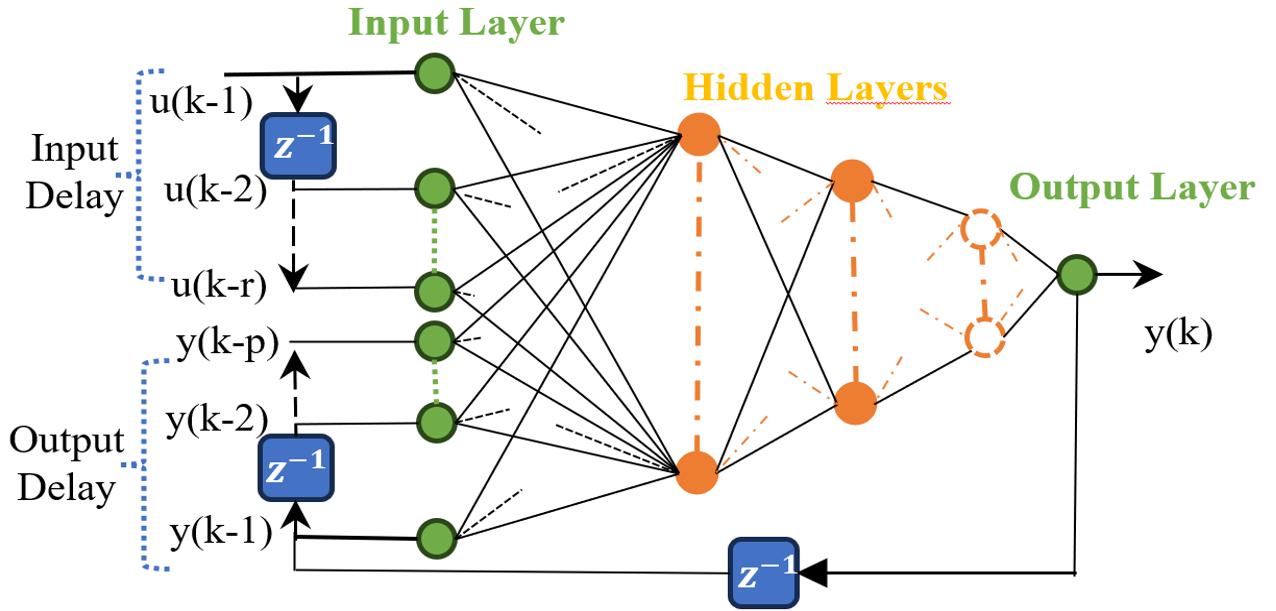}}
\caption{NARX NN overall structure.}
\label{fig2}
\end{figure}

Additionally, $e(k)$ signifies the modeling error or noise at sample $k$. $r$ and $p$ are integer positives that represent the input and output memories order, respectively. Fig. \ref{fig2} illustrates the standard architecture of the NARX NN featuring a single output.

\begin{figure}[htbp]
\centerline{\includegraphics [width=0.38\textwidth, height=0.12\textheight]
{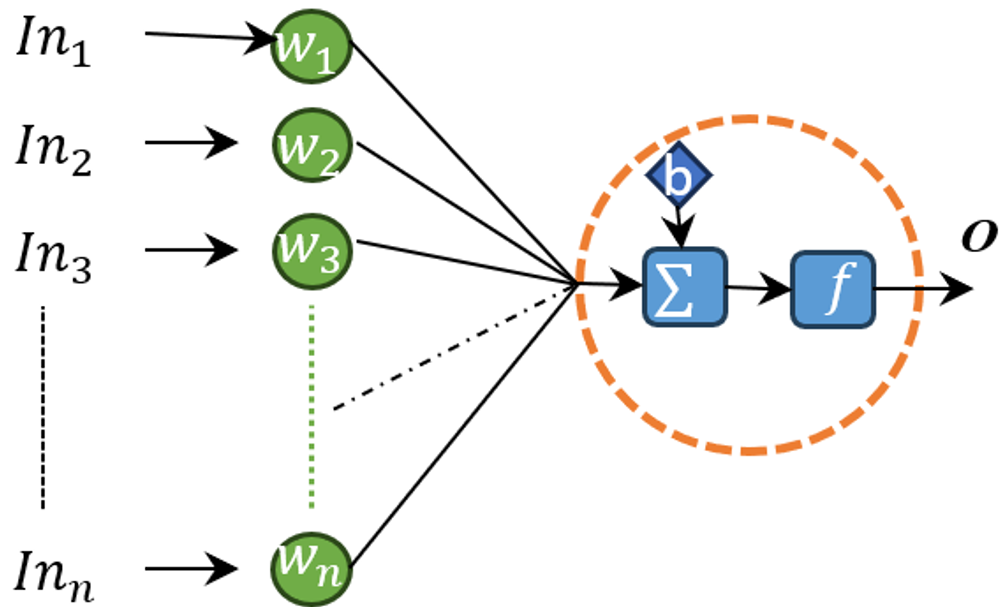}}
\caption{NN Neuron connection.}
\label{fig3}
\end{figure}

Fig. \ref{fig3} portrays the configuration of a NN neuron. 

The mathematical representation of Fig. \ref{fig3} is provided as follows: 

 \begin{equation}
 \label{eq2}
O=\mathrm{f}\left(\left(\sum_{i=1}^n \operatorname{In} i \times w i\right)+\mathrm{b}\right)
\end{equation}

In this context (\ref{eq2}), $n$ represents the number of input signals, $Ini$ stands for the input of the $i-th$ neuron, which is also dependent on the output of the $i-th$ neuron in the previous layer. $O$ denotes the neuron's output, $b$ refers to the weight bias, $w_i$ signifies the $i-th$ weight of the connection, all within the context of the activation function $f$.

The NARX NN, comprising a single hidden layer, is depicted as follows:
%\vspace{-3mm}

\begin{equation}
\label{eq3}
y(k)=f_{\text {o}}\left(W_{\text {o }}\left(f_h\left(W_h X(k-1)+b_{ h}\right)\right)+b_{\text {o }}\right)
\end{equation}

In (\ref{eq3}), the terms $y(k)$, $b_o$, $W_o$, and $f_o$ correspond to the approximated output, bias vector, weight matrix, and activation function of the output layer, respectively. Similarly, $b_h$, $W_h$, and $f_h$ denote the bias vector, weight matrix, and activation function of the hidden layer. Furthermore, within equation (\ref{eq3}), $X (k - 1)$ is expressed as:

%\vspace{-3mm}

\begin{equation}
\label{eq4}
\begin{aligned}
& X(k-1)=[y(k-1), y(k-2), \ldots, y(k-r), u(k- \\
& 1), u(k-2), \ldots, u(k-p)]^T
\end{aligned}
\end{equation}

\begin{figure}[htbp]
\includegraphics [width=0.38\textwidth, height=0.16\textheight]{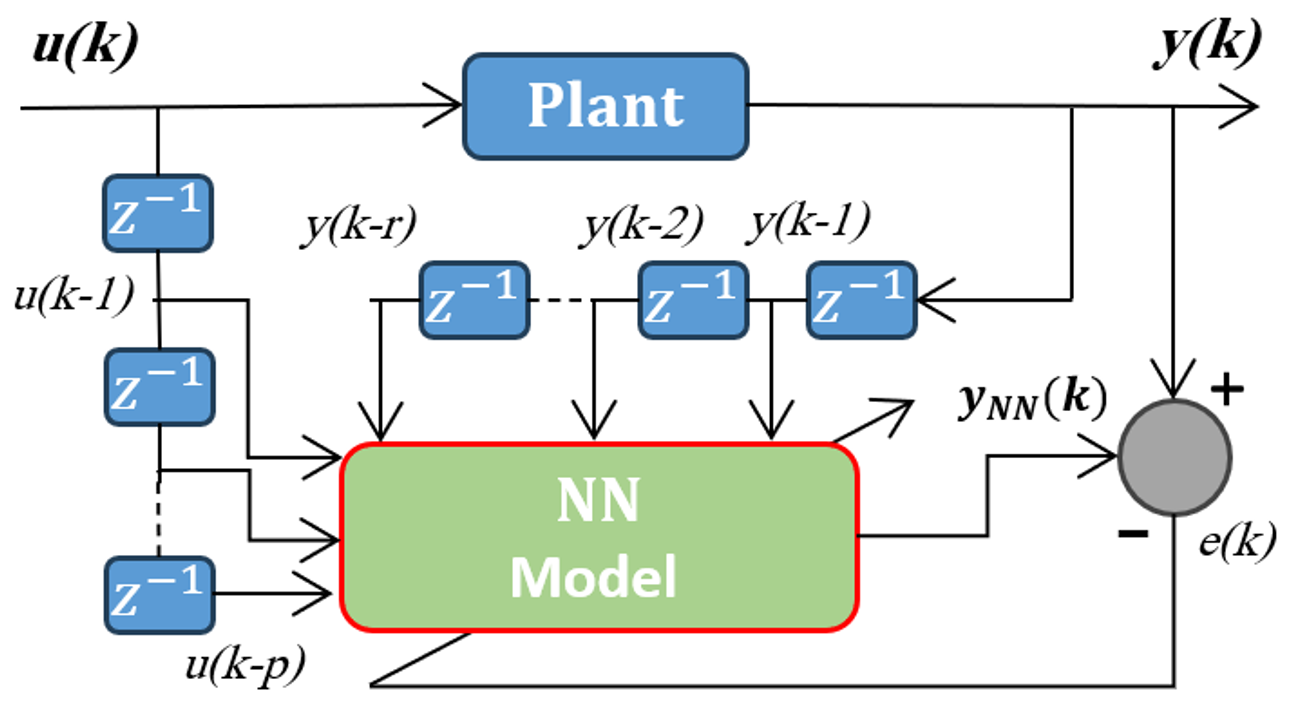}
\caption{NARX model overall Blocks.}
\label{fig4}
\end{figure}

During the offline training phase of a NARX NN, the objective is to optimize $b_o$, $W_o$, $b_h$, and $W_h$, thus creating a finely-tuned NN that functions effectively as an estimator. In the input data for the training phase of a NARX NN, there are two possible modes: parallel (P) mode and series-parallel (SP) mode \cite{15}. The distinction between these modes lies in the architecture: in the P mode, output feedback loops into the input, whereas in the SP mode, there is no such feedback loop. The general block diagram depicting the utilization of the NARX model with a plane is illustrated in Fig. \ref{fig4}.

In this paper, the series-parallel (SP) mode is utilized for both the training and operational phases of the NN. Real data values drive this choice during network operations. The NN employs a NARX model with a structure grounded in sigmoidal networks, featuring elements of a feedforward NN and a cascade NN.

\section{Results}
\subsection{Experimental Setup}
Due to the limited availability of information regarding quadrotor physical parameters, such as inertia moments and aerodynamic coefficients, we must adopt the Blackbox identification direct approach. In this approach, system identification is performed in the time domain, and each input channel of the quadrotor is excited individually. 

\begin{figure}[htbp]
\centerline{\includegraphics [width=0.40\textwidth, height=0.07\textheight]{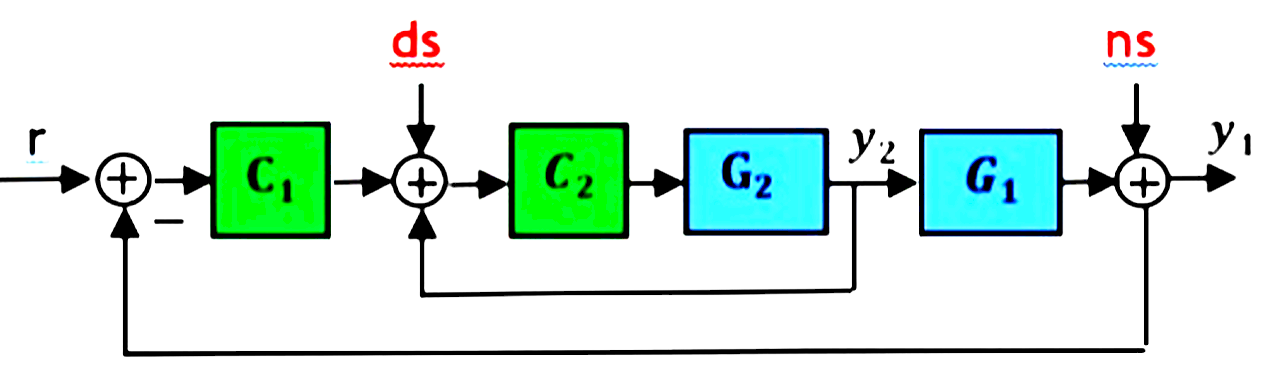}}
\caption{Quadrotor attitude dynamics under cascade controllers.}
\label{fig5}
\end{figure}

Given the inherent instability in quadrotor dynamics, we have implemented two feedback controllers to ensure the stability of attitude dynamics. Specifically, controller $C1$ stabilizes each angle of the quadrotor, while controller $C2$ focuses on stabilizing the angular velocities, as illustrated in Fig. \ref{fig5}. Here, $y1$ and $y2$ represent each channel's dynamic responses of angular velocity and angle, respectively. Additionally, $G1$ serves as a pure integrator, $r$ denotes the reference signal, $ds$ accounts for control disturbances, and $ns$ represents the output noise.
%\vspace{-3mm}

The target identification system $G2$ is subject to the sum of the well-selected PRBS signal \cite{3} and the output of angle rate controller C2. The response signal of the target dynamic $G2$   is collected for the identification process. 
  The primary controller, $C1$, tracks the angle reference. The gains for both controllers have been determined through trial and error and provide minimal angle and angular velocity stabilization. These controller settings have been carefully selected to maintain the desired levels of stabilization for both angle and angular velocity.
   In our time-domain identification approach, we have employed the PRBS as the input signal for our experimental tests. This signal serves as the excitation input for each attitude dynamic and has proven exceptionally effective when combined with time-domain identification methods. The PRBS is a periodically recurring deterministic signal that exhibits Gaussian noise-like characteristics \cite{3}.
Based on prior knowledge gained from analyzing quadrotor open-loop responses, it has been determined that the dominant dynamics typically occur within the range of 0.1 to 20 radians per second. Consequently, the PRBS excitation signal is crafted to align with this predefined range of interest \cite{3}.  
It's important to note that all outputs of the quadrotor model involve integrators. These integrators are numerically unstable and cannot be reliably identified. To address this issue, our modeling approach utilizes the ($U_1, U_2,U_3,U_4$) control signals to directly measure the attitude derivatives outputs ($p, q, r, z$) from inertial sensors. Specifically, $p$ is approximated as 
$p \approx \dot{\phi}, q \approx \dot{\theta}, r \approx \dot{\psi}$, roll pitch and yaw angle-rats of the quadrotor, while angles near operating point. To model the complex relationships governing the quadrotor's attitude, we employed a NARX neuronal network model. This model excels in capturing nonlinear and time-delayed dependencies in such systems, especially when they deviate significantly from the operating point, exceeding 20°. We designed and trained separate NARX NN architectures for each roll, pitch, and yaw rate, tailoring the network configurations to the unique characteristics of each rate.

\subsection{Validation}

In this paper, we employ NARX-NN to identify the dynamic attitude of our quadrotor across various configurations as follows:

\paragraph{NARX Model With A Sigmoid NN} 

\begin{figure}[htbp]
\centerline{\includegraphics [width=0.42\textwidth, height=0.2\textheight]{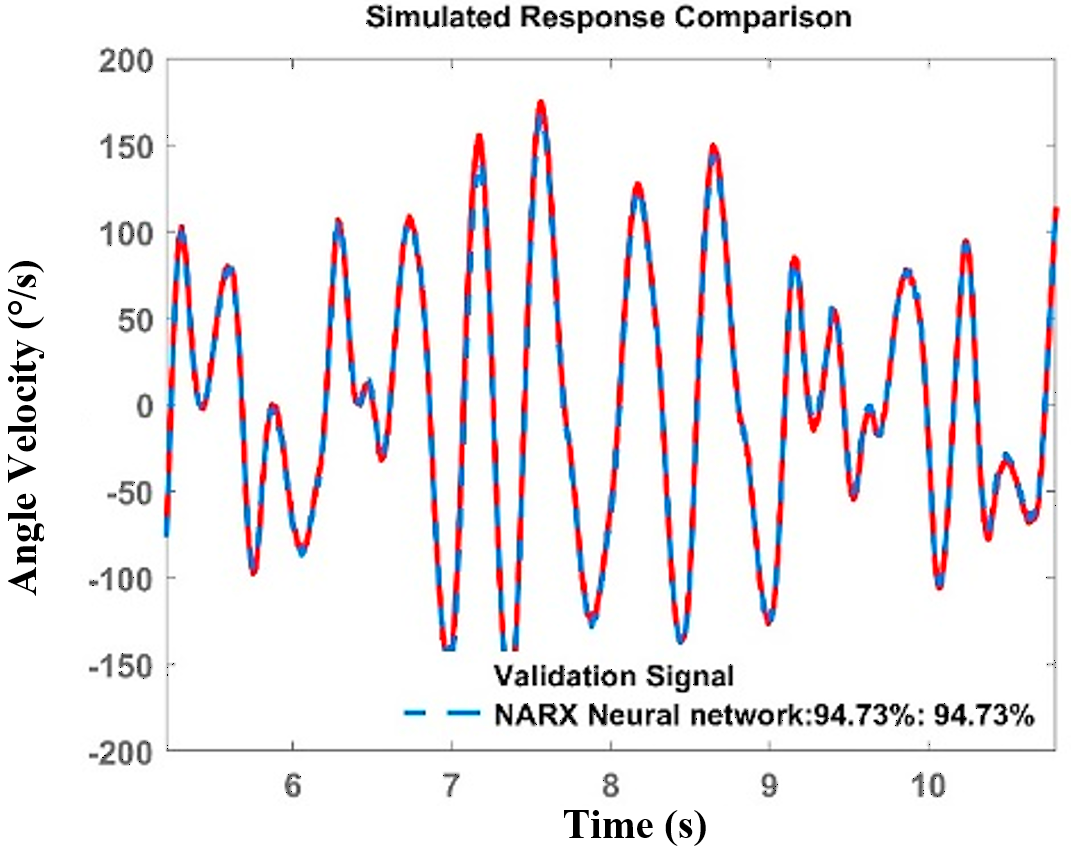}}
\caption{Evaluation of NARX Sigmoid NN Model.}
\label{fig6}
\end{figure}

Here, we employed a NARX model with a sigmoid NN component to identify the dynamic behavior of a quadrotor's angle rate. It specifies a model with 15 output delays, 7 input delays, and 0 noise delays. We create a NN model with a sigmoid activation function. It has 30 hidden units in its single hidden layer. This NN is used as a nonlinear function approximator within the NARX model. The trained model demonstrated exceptional predictive accuracy, achieving a fit of 99.26\% on the estimation data and 94.73\% on the validation data. Key performance metrics, such as Final Prediction Error (FPE) and Mean Squared Error (MSE), further validated the model's effectiveness, with low values of 1.236 and 0.3266, respectively. This indicates that the model not only accurately captured the underlying dynamics but also demonstrated strong generalization capabilities when applied to unseen data, as illustrated in Fig. \ref{fig6}.
%\vspace{-3mm}

\paragraph{NARX Feedforward NN}
Here, a nonlinear NARX model was employed, incorporating a feedforward NN (FFNN) with two hidden layers of 10 and 20 neurons for the first and second hidden layer to be used as a nonlinear function approximator within the NARX model, where the radial basis set for second hiding layer. The NARX model was trained using time-domain data related to quadrotor angles-rate dynamics. During validation, the model exhibited outstanding performance, with a fit of 99.67\% on the estimation data and a MSE of 0.06726, reflecting precise capturing of the system's behavior. Furthermore, when applied to validation data, the model maintained its high accuracy, achieving a fit of 95.95\%. These results underscore the model's robustness and its ability to generalize effectively to unseen data, confirming its suitability for dynamic system identification in this context as Fig. 7.
Our analysis revealed a gradual decrease in the training MSE, indicating improved fitting to the training data over epochs Fig. 8. Importantly, we observed a notable milestone at epoch 50, where the validation performance reached its peak with an MSE of 0.091288. This signifies the model's optimal generalization to unseen data, validating its reliability for real-world applications. Although not explicitly shown, we also conducted rigorous testing on independent data to confirm the model's robustness.
%\vspace{-2mm}

\begin{figure}[htbp]
\centerline{\includegraphics [width=0.42\textwidth, height=0.22\textheight]{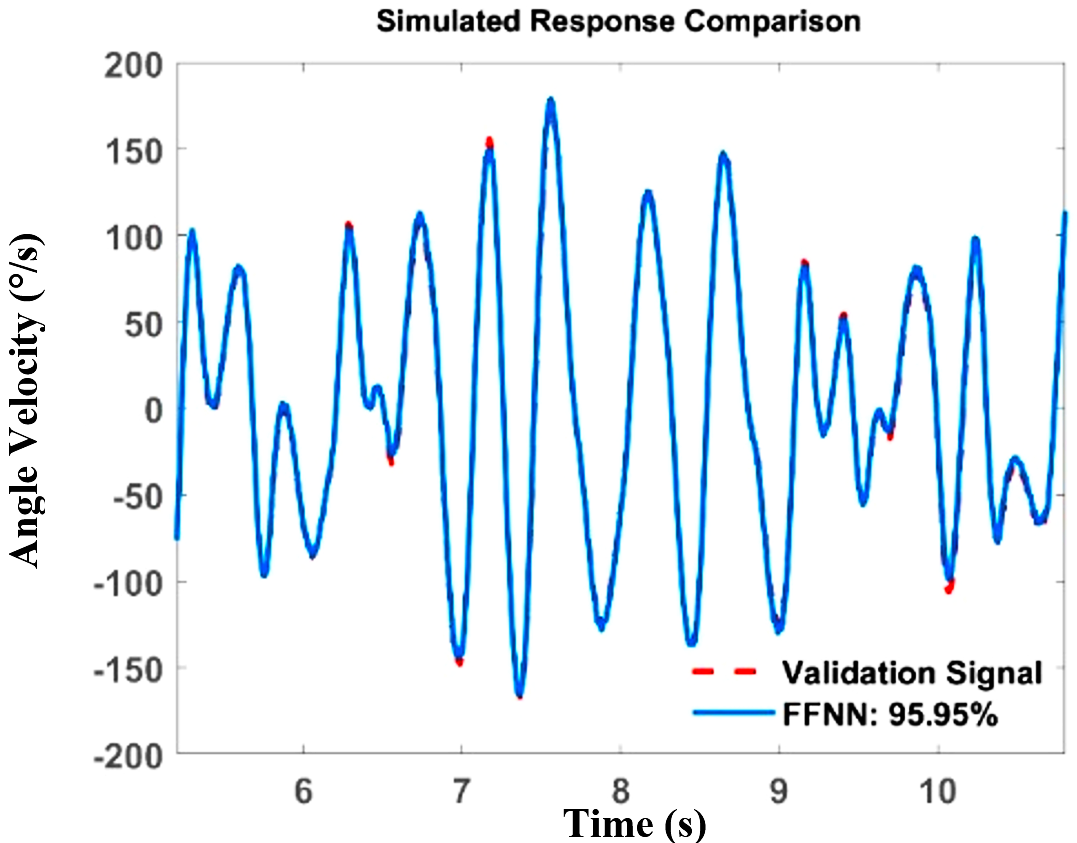}}
\caption{Evaluation of NARX Feedforward NN.}
\label{fig7}
\end{figure}

This comprehensive evaluation reaffirms the effectiveness and practicality of our trained NN model in the context of our research.

\begin{figure}[htbp]
\centerline{\includegraphics [width=0.45\textwidth, height=0.25\textheight]{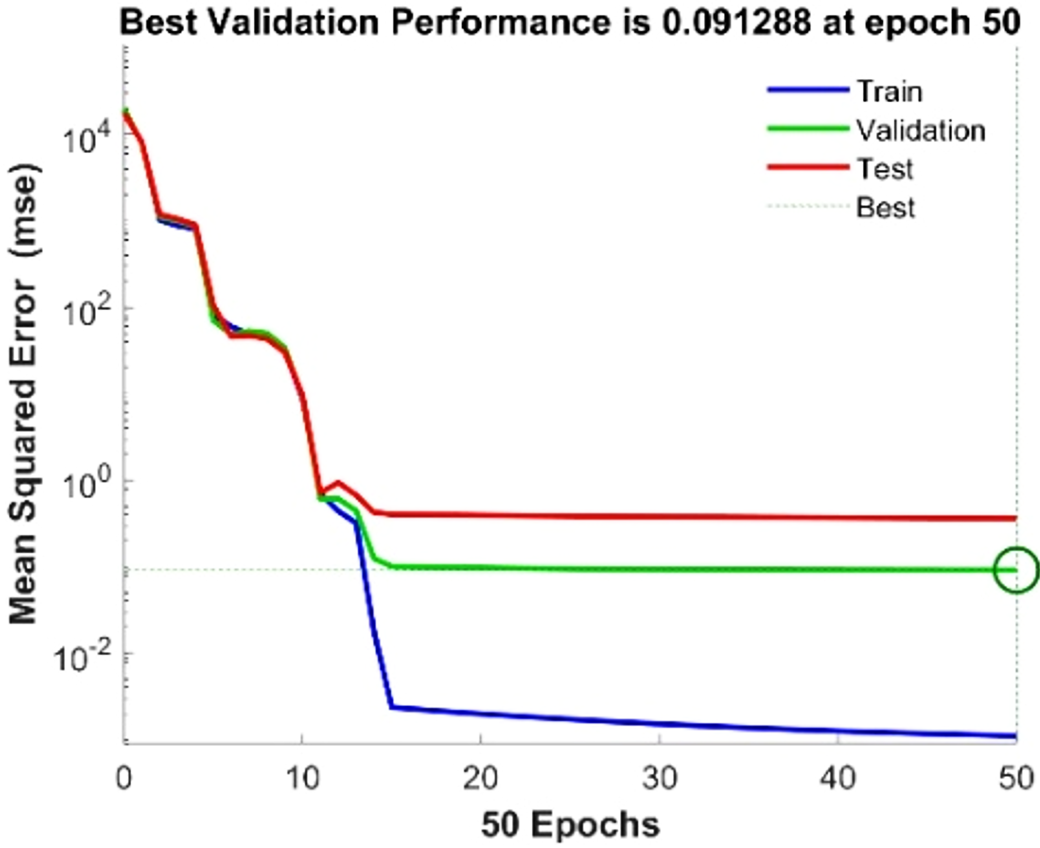}}
\caption{NARX Feedforward NN performance }
\label{fig8}
\end{figure}

\begin{figure}[htbp]
\centerline{\includegraphics [width=0.50\textwidth, height=0.4\textheight]{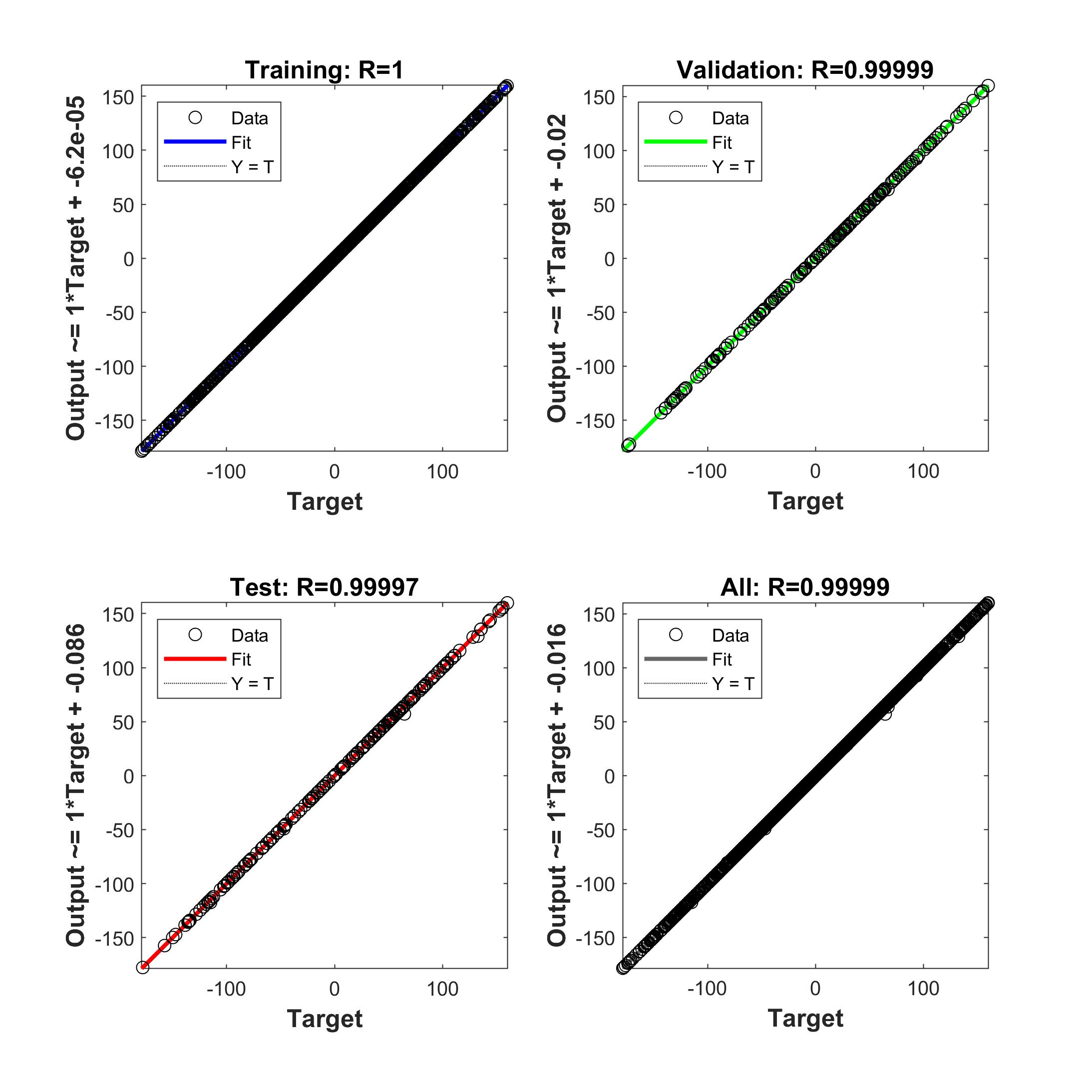}}
\caption{Correlation of NN Predictions with collected signal.}
\label{fig9}
\end{figure}
%\vspace{-4mm}

a visual representation in Fig. 9 of how well our NN's predictions align with the actual target values in a regression task, our NN model exhibited a remarkable correlation of 1 between the predicted values and the actual target values in the training set and 0.99997 in validation set. 
This perfect alignment is indicative of the model's ability to precisely capture the underlying patterns in the training data. The linear regression line closely follows the diagonal line, further confirming the accuracy of our model's predictions.

\paragraph{Cascade NN} 

Here, we employed a NARX model with a cascade forward NN as the estimator to identify quadrotor angle-rate dynamics. cascade forward NN is created with a hidden layer containing 20 neurons. The NARX model exhibited outstanding performance, achieving a fit of 99.65\% on the training data. With the accuracy of a predictive model MSE 0.07344, which indicates that the model's predictions closely match the actual data points during training. And 98.47\% on the validation data. These results as Fig.10 demonstrate the model's ability to accurately capture and predict system behavior, underscoring its effectiveness for dynamic system identification. 

\begin{figure}[htbp]
\centerline{\includegraphics [width=0.42\textwidth, height=0.22\textheight]{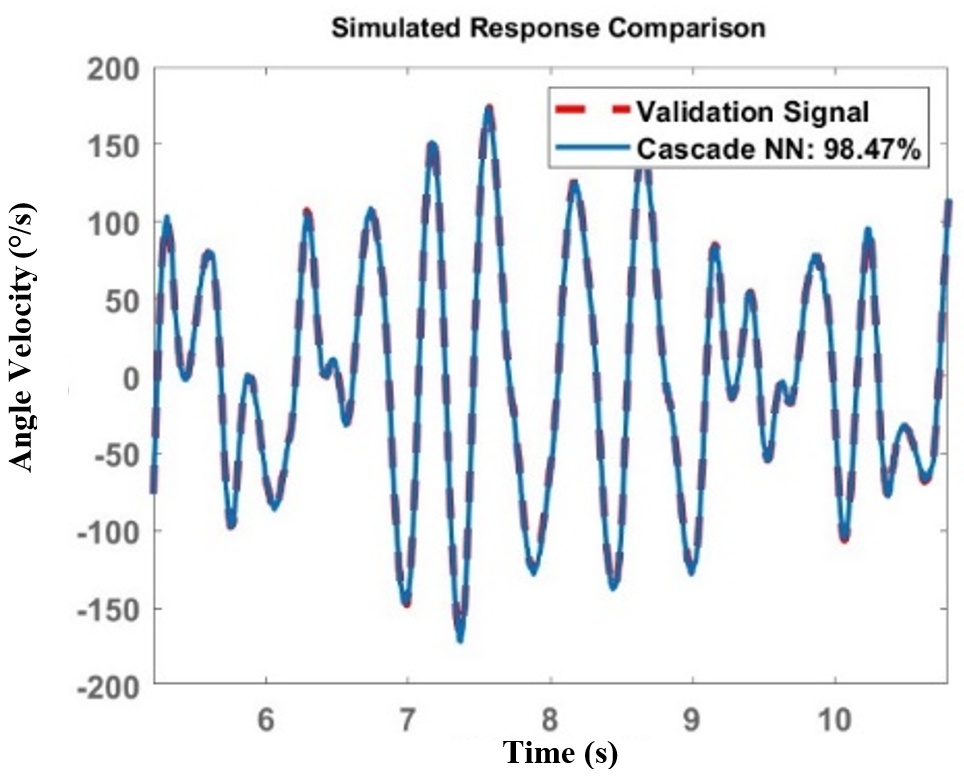}}
\caption{Evaluation of Cascade NN Model.}
\label{fig10}
\end{figure}

\section{CONCLUSION}
In summary, our research has highlighted the significant contributions of NARX-NN to the field of quadrotor dynamics modeling and system identification. Through a comprehensive analysis and rigorous experimentation, we have shown that these networks excel in capturing the intricate and nonlinear behaviors inherent in quadrotor dynamics. Our models consistently achieved high levels of accuracy and robustness, as evidenced by fit percentages exceeding 99\% on estimation and validation data. Furthermore, the comparative study conducted in this research underscores the unique advantages of NARX networks when compared to other NN architectures. The NARX model with its sigmoid and feedforward NN, as well as the cascade NN, outperformed alternative approaches in accurately representing quadrotor dynamics. The potential applications of our findings are far-reaching, extending to fields such as aerial robotics, autonomous navigation, and beyond. By providing accurate and comprehensive modeling capabilities, NARX-NN have the capacity to enhance quadrotor control, enabling these versatile systems to perform complex tasks in real-world environments with greater precision. Future studies will focus on the design and implementation of artificial intelligence controllers, such as the Takagi-Sugeno fuzzy, neuronal network, etc., in low-cost microcontrollers like our STM32 in order to stabilize the quadcopter during trajectory tracking.

\section*{Acknowledgment}

This work was supported by the Laboratory of Energetic System Modelling (LMSE) of the University of Biskra, Algeria, under the patronage of the General Directorate of Scientific Research and Technological Development (DGRSDT) in Algeria. The research project was approved by the Ministry of Higher Education and Scientific Research in Algeria, under the number A01L08UN070120220003.


\end{document}